\documentstyle[prd,aps,floats,tighten,epsfig]{revtex}

\begin{document}
\draft \twocolumn

%\widetext
\title{On the Domain of Mixing Angles in Three Flavor Neutrino Oscillations}
\author{Jonny Lundell\footnotemark[1] and H{\aa}kan Snellman\footnotemark[2]} \address{Theoretical Physics, Department of
Physics, Royal Institute of Technology,\\ SE-100 44 STOCKHOLM, SWEDEN}
\date{\today}

\maketitle

\def\thefootnote{\fnsymbol{footnote}}
\footnotetext[1]{Electronic address: jonny@theophys.kth.se}
\footnotetext[2]{Electronic address: snell@theophys.kth.se}

\begin{abstract}
We clarify the domain needed for the mixing angles in three flavor
neutrino oscillations. By comparing the ranges of the transition
probabilities as functions of the domains of the mixing angles, we show
that it is necessary and sufficient to let
all mixing angles be in $\left[ 0, \pi/2 \right]$. This holds
irrespectively of any
assumptions on the neutrino mass squared differences. 
\end{abstract}


\section{Introduction}
\label{partdomain}

With the growing realization that neutrinos might have non-zero masses, studies
of neutrino
physics data in terms of neutrino
oscillations, with three (or more) flavors, are becoming more and more popular.

The easiest way to describe the neutrino oscillations is to use a set
of two
flavor oscillation schemes. However, since we observe three neutrino
flavors in nature, a
more natural approach is to use a three flavor oscillation model, and there are
reports of different results obtained, when using two or three flavors
\cite{ThunMcKeeOhlssonSnellman,BarenboimScheckSakaiTeshima}.

Much work has been done on neutrino oscillations using two flavor
models. When using a three flavor oscillation model,
one simplification often used is to make assumptions of the mass
squared differences to enable the problem to be treated using much
of the results coming from the use of two flavor oscillation models.

In a two flavor oscillation model it is natural to let the mixing
angles be in $\left [ 0,\pi/4 \right ]$.
By assuming the mass squared differences to be in certain ranges that
allow the three flavor oscillation model to be simplified, using
results from the two flavor oscillation model, this domain of the
mixing angles is often inherited to the three flavor oscillation models.
Several authors that have made such assumptions about the mass squared differences
have restricted the mixing angles to $[0,\pi/4]$ \cite{FogliLisiMontaninoconfortokim}.
Other authors, however, have found solutions
with mixing angles larger than $\pi/4$ when using a
three flavor oscillation model \cite{BarenboimScheckSakaiTeshima}.

To clarify the situation we here perform a detailed investigation of the domain
needed for the vacuum mixing angles without any specific assumptions and
show that for a correct analysis to be possible, it is necessary to let the
domain be $0 \leq \theta_{i} \leq \pi/2$ for all $i=1,2,3$ \cite{HarariLeurer}. This holds
even if assumptions on the mass squared differences are made. 

Our paper is organized as follows. In Section 2 we describe our
formalism. In Section 3 we analyze the domain needed for three flavor
oscillations, and in Section 4 we discuss the results.

\section{Formalism}
\label{sectionformalism}

For the present analysis we will use the plane wave approximation when
describing the neutrino propagation. We will describe the neutrino flavor
states produced by charge
current weak interactions as a mixing of neutrino mass eigenstates. For
simplicity we neglect any $CP$-changing phases.  We then write the flavor
states \mbox{$\vert \nu_\alpha \rangle,\;\alpha=e, \mu, \tau$}, as
a linear combination of the mass eigenstates
\mbox{$\vert \nu_i \rangle,\; i=1,2,3$},
\begin{equation}
\vert \nu_\alpha \rangle = \sum^3_{i=1} U^*_{\alpha i} \vert \nu_i
\rangle , \hspace{1cm} \alpha=e, \mu, \tau.
\end{equation}
Here $U$ is a unitary mixing matrix, given by the standard
representation of the Cabibbo-Kobayashi-Maskawa mixing matrix as
\begin{equation}
U = \left( \begin{array}{ccc} C_2 C_3 & S_3 C_2 & S_2 \\ - S_3 C_1 -
S_1 S_2 C_3 & C_1 C_3 - S_1 S_2 S_3 & S_1 C_2 \\ S_1 S_3 - S_2
C_1 C_3 & - S_1 C_3 - S_2 S_3 C_1 & C_1 C_2 \end{array} \right),
\end{equation}
with $C_i\equiv\cos{\theta_i}$ and $S_i\equiv\sin{\theta_i}$, $i=1,2,3$. Since
there are no $CP$ phases, the matrix elements are real and we have
$U^\ast_{\alpha
a} = U_{\alpha a}$ for $\alpha = e, \mu, \tau$ and $a = 1,2,3$.

The transition probability for a neutrino changing from $\nu_\alpha$
to $\nu_\beta$ is given by
\begin{equation}
P_{\alpha \beta}= \delta_{\alpha \beta}-4\mathop{\sum_{i=1}^3 \sum_{j=1}^
3}_{i<j} U_{\alpha i} U_{\beta i} U_{\alpha j} U_{\beta j} \sin^
2 \frac{\Delta m^2_{ij} L}{4E},
\label{Palphabeta}
\end{equation}
where $\alpha,\beta=e, \mu,\tau$. Here $\Delta m^2_{ij}\equiv m^2_i-m^2_j$,
$E$ is the relativistic
energy of the neutrinos and $L$ is the source-detector distance.

The mass squared differences are not all independent in a three neutrino
model, but fulfill by definition the relation
\begin{equation}
\label{deltam}
\Delta m^2_{21}+\Delta m^2_{32}+\Delta m^2_{13}=0.
\end{equation}
From Eq. (\ref{Palphabeta}) it follows that $P_{\alpha \beta} =
P_{\beta \alpha}$.

In what follows we shall, without loss of generality, assume that the
masses are ordered so that $m_{1} \leq m_{2} \leq m_{3}$, and
the mass squared differences $\Delta m^2_{ij}$ are non-negative when
$i > j$.

From the unitarity condition we have three equations that the
transition probabilities must fulfill:
\begin{eqnarray}
\label{unitary1}
P_{e e} + P_{e \mu} + P_{e \tau} &=&1,\\
P_{e \mu} + P_{\mu \mu} + P_{\mu \tau} &=&1,\\
\label{unitary3}
P_{e \tau} + P_{\mu \tau} + P_{\tau \tau} &=&1.
\end{eqnarray}
Since we want to investigate the mixing angles $\theta_i$, we will for
convenience introduce the auxiliary `angles'
\begin{eqnarray}
\Phi_1=\frac{\Delta m^2_{21} L}{4 E}\hspace{3mm}{\rm and}\hspace{3mm}
\Phi_2=\frac{\Delta m^2_{32} L}{4 E}.
\label{phi1phi2}
\end{eqnarray}
Using Eq. (\ref{deltam}), this gives
\begin{equation}
\frac{\Delta m^2_{31} L}{4 E}=\Phi_1 + \Phi_2.
\end{equation}
The auxiliary angles are both positive by definition.
To introduce the auxiliary angles makes sense, because we do not here care
about the
mass squared differences. The values of $\Phi_1$ and $\Phi_2$ vary with
$L/E$, which is something that is determined by the experimental
setup. The mass squared differences only determine how fast the
auxiliary angles vary.

We can now write $P_{\alpha \beta}$ as
\begin{eqnarray}
P_{\alpha \beta}&=&\delta_{\alpha \beta}  - 4 \left[ f^1_{\alpha \beta}
 \sin^2 \Phi_1  + f^2_{\alpha \beta} \sin^2 \Phi_2 \right. \nonumber\\
 & & + \left. f^3_{\alpha \beta} \sin^2 (\Phi_1 + \Phi_2) \right],
\end{eqnarray}
where $f^k_{\alpha \beta}=f^k_{\alpha \beta}(\theta_1,\theta_2,\theta_3)$
is given by
\begin{eqnarray}
f^1_{\alpha \beta}&=&U_{\alpha 1} U_{\beta 1} U_{\alpha 2} U_{\beta
2},\nonumber\\
f^2_{\alpha \beta}&=&U_{\alpha 2} U_{\beta 2} U_{\alpha 3} U_{\beta
3},\nonumber\\
f^3_{\alpha \beta}&=&U_{\alpha 1} U_{\beta 1} U_{\alpha 3} U_{\beta 3}.
\end{eqnarray}

\section{Analysis}
\label{sectionanalysis}

To determine the necessary and sufficient domain for $\theta_i,\,
i=1,2,3,$ we first  show
that it is sufficient to take $\theta_i \in [0,\pi/2]$, and then continue
to look whether it is possible to restrict the domain further.

We first notice, that due to unitary, it is sufficient to look at the diagonal
probabilities, $P_{e e}$, $P_{\mu \mu}$, and $P_{\tau \tau}$. The 
nondiagonal probabilities $P_{e \mu}$, $P_{e \tau}$, and $P_{\mu \tau}$,
can be calculated from the diagonal ones using Eqs.
(\ref{unitary1})-(\ref{unitary3}).

The diagonal probabilities are given by
\begin{equation}
P_{\alpha \alpha}= 1-4\mathop{\sum_{i=1}^3 \sum_{j=1}^
3}_{i<j} U_{\alpha i}^2 U_{\alpha j}^2 \sin^2 \frac{\Delta m^2_{ij}
L}{4E}, \hspace{5mm} \alpha=e, \mu, \tau.
\end{equation}

Since the $U_{\alpha i}U_{\alpha j}$ appear everywhere as $U_{\alpha
i}^2U_{\alpha j}^2$, every $\sin
\theta_i$ and $\cos \theta_i$ will appear as squared, causing  the
signs to
disappear. It is therefore always possible to choose an angle between $0$
and $\pi/2$
to parameterize the mixing angles. A sufficient domain for their
variation is therefore given by $[0,\pi/2]$.

In order to see if it is possible to reduce the domain for $\theta_i$ further, we want to find a
way to handle the problem that the $P_{\alpha \alpha}$'s are functions of
five variables, two mass squared differences and three mixing angles. We
will do that
by fixing $\Phi_1$ and $\Phi_2$. For every fixed  $\Phi_1$ and
$\Phi_2$ we then have the $P_{\alpha \alpha}$'s as functions of the three
mixing angles. This means that we have solved the problem of having too
many parameters at the cost of getting a lot of functions instead. For
each pair of $\Phi_1$ and $\Phi_2$ there is a function that depends on three
parameters, the mixing angles.

Next, it turns out to be practical not to explicitly look for the domain
needed for the
mixing angles, but instead check if there is any difference between different domains of the angles. If we look at one of the
$P_{\alpha \alpha}$'s, it
will take some range of values for some fixed domain of values of the
$\theta_i$'s. If
there are some values it cannot take when the angles are restricted to a smaller
domain, the restriction makes a difference. We will therefore look at the range
of $P_{\alpha \alpha}$, denoted $R(P_{\alpha \alpha})$.

Considering one probability at a time and without any
restrictions on the parameters, the range of $P_{\alpha \alpha}$ is of course
$R(P_{\alpha \alpha})=[0,1]$.
However, for fixed values of $\Phi_1$ and $\Phi_2$, the range
$R(P_{\alpha \alpha})$ can be smaller. To
see this, take $L=0$, giving $\Phi_1=\Phi_2=0$, as an example. This
leads to $P_{\alpha \alpha}=1$ for all possible
$\theta_i$'s, which is obviously a smaller range.

To simplify the forthcoming discussion we introduce
\begin{eqnarray}
P'_{\alpha \alpha}&=& 4 \left[ f^1_{\alpha \alpha}  \sin^2 \Phi_1  +
f^2_{\alpha \alpha} \sin^2 \Phi_2 \right. \nonumber\\
& & +\left. f^3_{\alpha \alpha} \sin^2 (\Phi_1 + \Phi_2) \right] .
\end{eqnarray}
We do this because all $f^k_{\alpha
\alpha},\;k=1,2,3,$ have at least one factor $\sin \theta_i$ for some $i$, which
means that
\begin{equation}
P'_{\alpha \alpha}=0 \hspace{5mm} {\rm when} \hspace{5mm}
\theta_1=\theta_2=\theta_3=0,
\end{equation}
and we thus obtain a range of the type
$[0,x]$, where $x \leq 1,$ for all $P'_{\alpha \alpha}$'s.
We see that the interesting value is the upper limit of $R(P'_{\alpha
\alpha})$.

From now on we will always consider $\Phi_1$ and $\Phi_2$ as
fixed when calculating $R(P'_{\alpha \alpha})$, making $R(P'_{\alpha
\alpha})$ a function of $\Phi_1$ and $\Phi_2$. This means that we take
fixed $\Phi_1$ and $\Phi_2$ and then calculate the range of the
function, which now has the three mixing angles as parameters. Each
choice of $\Phi_1$ and $\Phi_2$ will give a specific range. The upper limit of
the range can obviously change for different domains of the
$\theta_i$'s. Introduce the notation

\begin{eqnarray}
R(P'_{\alpha \alpha})_{abc} \equiv \left \{\right. P'_{\alpha
\alpha}(\theta_1,\theta_2,\theta_3) :\; \theta_1 \in [0,a],\nonumber\\
\theta_2 \in [0,b],\,\theta_3 \in [0,c] \left.\right \}
\end{eqnarray}

\begin{figure}[t]
\centering
\begin{picture}(220,186)(-7,-10)
\put(0,0){\epsfig{figure=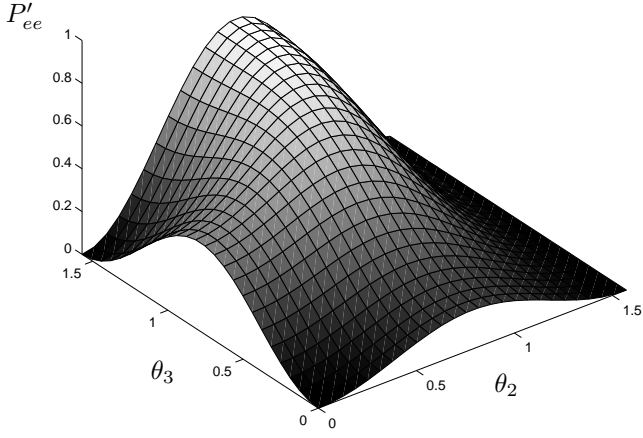,clip=,width=80mm}}
\put(40,20){$\theta_{3}$}
\put(170,15){$\theta_{2}$}
\put(-15,155){$P'_{ee}$}
\end{picture}
\caption{ $P'_{e e}$, $\Phi_1=0.7$ and $\Phi_2=2.0$. Note that $P'_{e
e}$ does not depend on $\theta_1$.}
\label{p11thetaplot}
\end{figure}
\begin{figure}[!b]
\centering
\begin{picture}(220,175)(-10,-10)
\put(0,0){\epsfig{figure=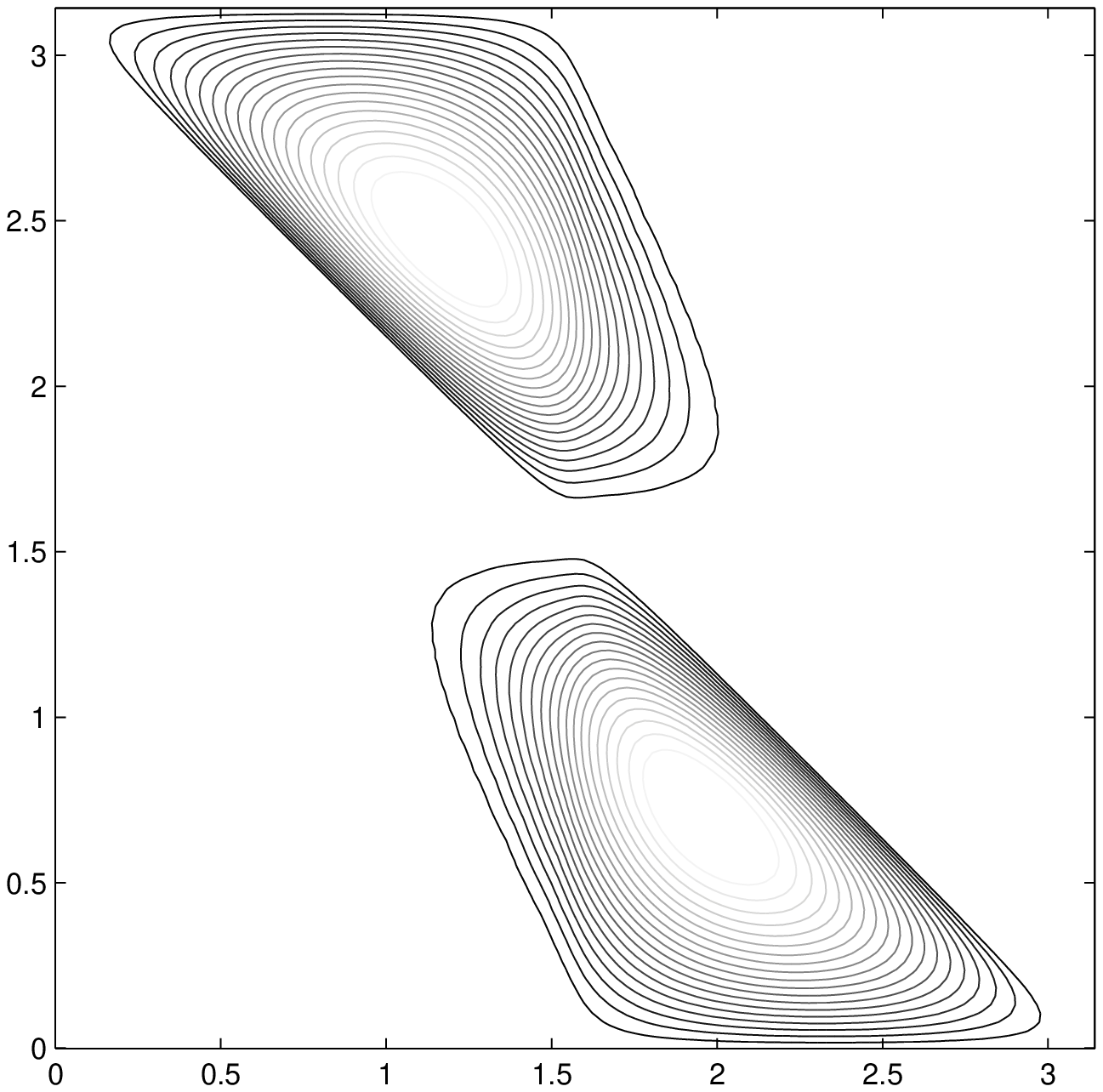,height=160pt}}
\put(80,-8){$\Phi_2$}
\put(-10,80){\rotatebox{90}{$\Phi_1$}}
\put(80,165){$P_{e e}$}
\end{picture}
\caption{Plot of $R(P'_{e
e})_{\frac{\pi}{2}\frac{\pi}{2}\frac{\pi}{2}} - R(P'_{e
e})_{\frac{\pi}{4} \frac{\pi}{4} \frac{\pi}{4}}$.}
\label{p11phiplot}
\end{figure}
\begin{figure}[!t]
\centering
\begin{picture}(220,175)(-10,-10)
\put(0,0){\epsfig{figure=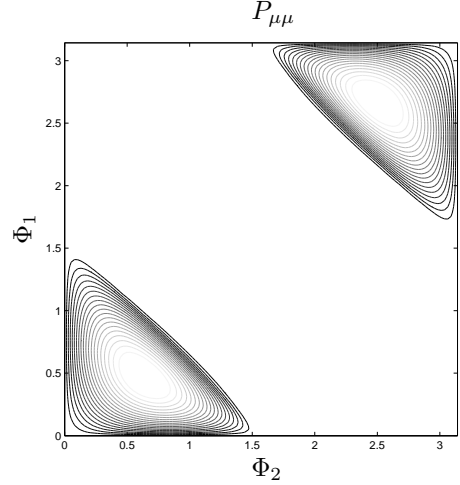,height=160pt}}
\put(80,-8){$\Phi_2$}
\put(-10,80){\rotatebox{90}{$\Phi_1$}}
\put(80,165){$P_{\mu\mu}$}
\end{picture}
\caption{Plot of $R(P'_{\mu
\mu})_{\frac{\pi}{2}\frac{\pi}{2}\frac{\pi}{2}} - R(P'_{\mu
\mu})_{\frac{\pi}{4} \frac{\pi}{4} \frac{\pi}{4}}$.}
\end{figure}
\begin{figure}[!b]
\centering
\begin{picture}(220,175)(-10,-10)
\put(0,0){\epsfig{figure=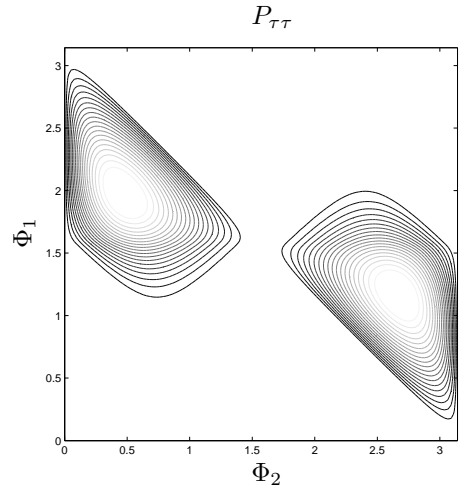,height=160pt}}
\put(80,-8){$\Phi_2$}
\put(-10,80){\rotatebox{90}{$\Phi_1$}}
\put(80,165){$P_{\tau\tau}$}
\end{picture}
\caption{Plot of $R(P'_{\tau
\tau})_{\frac{\pi}{2}\frac{\pi}{2}\frac{\pi}{2}} - R(P'_{\tau
\tau})_{\frac{\pi}{4} \frac{\pi}{4} \frac{\pi}{4}}$.}
\label{p33phiplot}
\end{figure}

We now calculate $R(P'_{\alpha \alpha})_{abc}$,
$abc=\frac{\pi}{2}\frac{\pi}{2}\frac{\pi}{2}$,
$\frac{\pi}{4}\frac{\pi}{4}\frac{\pi}{4}$. As the
$f_k^{\alpha \beta}$'s are `nice' functions, that vary slowly, this is not
a problem to do numerically. Fig.
\ref{p11thetaplot} shows an example of how smoothly $P'_{e e}$ varies with
$\theta_i$.

The result is, that for some values of $\Phi_1$ and $\Phi_2$ we have
\begin{eqnarray}
R(P'_{\alpha \alpha})_{\frac{\pi}{2}\frac{\pi}{2}\frac{\pi}{2}}
\neq R(P'_{\alpha \alpha})_{\frac{\pi}{4}\frac{\pi}{4}\frac{\pi}{4}}. 
\end{eqnarray}

This means that for some values of $\Phi_1$ and $\Phi_2$, there are
values of the functions $P'_{\alpha \alpha}$ that can be obtained by
letting $\theta_i \in [0,\pi/2]$, but not be when $\theta_i \in
[0,\pi/4]$. 

In the same way, one can check what happens when only one or two mixing
angles are restricted to $[0,\pi/4]$. The result is that it is
possible to restrict either $\theta_1$ or $\theta_2$, but not both, to the smaller
domain without getting any differences in the ranges of the $P_{\alpha \alpha}$'s.

To visualize this result we define the difference of $R(P'_{abc})$ and
$R(P'_{a'b'c'})$ as the difference of the upper limit of the ranges
\begin{eqnarray}
&&R(P'_{\alpha \alpha})_{abc} - R(P'_{\alpha \alpha})_{a'b'c'}\nonumber\\
&&= \max_{\theta_{1,2,3}<a,b,c} \left (P'_{\alpha \alpha} \right ) -
\max_{\theta_{1,2,3} < a',b',c'} \left (P'_{\alpha \alpha} \right ).
\end{eqnarray}

We can now make a contour plot of this difference
with $\Phi_1$ and $\Phi_2$ on the axes. The result is displayed in
Figs.\ref{p11phiplot}-\ref{p33phiplot}.
In the empty areas of the Figs. there is no difference between the
domains, but inside each curve there is a difference that gets larger
as we move inwards. The plots have a two-fold rotational symmetry around
the point $\Phi_{1}=\Phi_{2}=\pi/2$.

For example, in $R(P'_{e e})$ we see from Fig. \ref{p11phiplot}, that the
maximum difference occurs when $\Phi_1 \simeq 0.7$ and $\Phi_2 \simeq
2$, (or $\Phi_1 \simeq 2.4$ and $\Phi_2 \simeq 1.1$). 

If we look at the probability at that point, we have
\begin{eqnarray}
P'_{e e}&=&4 \left [\right .(C_2 C_3 S_3 C_2)^2 \sin^2 \Phi_1
 + (S_3 C_2 S_2)^2 \sin^2 \Phi_2 \nonumber\\
 & &+ \left .(C_2 C_3 S_2)^2 \sin^2 (\Phi_1 + \Phi_2) \right ].
\end{eqnarray}
Since this equation only depends on $\theta_2$ and $\theta_3$, its range
can easily
be shown in the plot displayed in Fig. \ref{p11thetaplot}. It is clear that
the range is larger when allowing $\theta_3 > \pi/4$.

To see how large the difference is, we have given the maximum difference for
each of the $P'_{\alpha \alpha}$  in Table \ref{differencetable}. It
can readily be seen that the maximum differences are large enough to be of
importance.

\begin{table}[!h]
\begin{tabular}{@{\hspace{1.5 cm}} l  c @{\hspace{1.5 cm}} }
$P_{\alpha \alpha}$ & Maximum difference\\
\hline\hline$P_{e e}$ 	& $\approx 0.19$\\
\hline$P_{\mu \mu}$ 	& $\approx 0.05$\\
\hline$P_{\tau \tau}$ 	& $\approx 0.19$\\
%\hline$P_{e \mu}$ 	& 0.69\\
%\hline$P_{e \tau}$ 	& 0.50\\
%\hline$P_{\mu \tau}$ 	& 0.60\\
\end{tabular}
\caption{Maximum difference between  $R(P'_{\alpha
\alpha})_{\frac{\pi}{2}\frac{\pi}{2}\frac{\pi}{2}}$ and
$R(P'_{\alpha \alpha})_{\frac{\pi}{4}\frac{\pi}{4}\frac{\pi}{4}}$.}
\label{differencetable}
\end{table}

For the nondiagonal probabilities the differences are even larger.

\section{Discussion and Conclusion}
\label{sectiondiscussion}
\begin{figure}[!b]
\centering
\begin{picture}(220,180)(-10,-20)
\put(0,0){\epsfig{figure=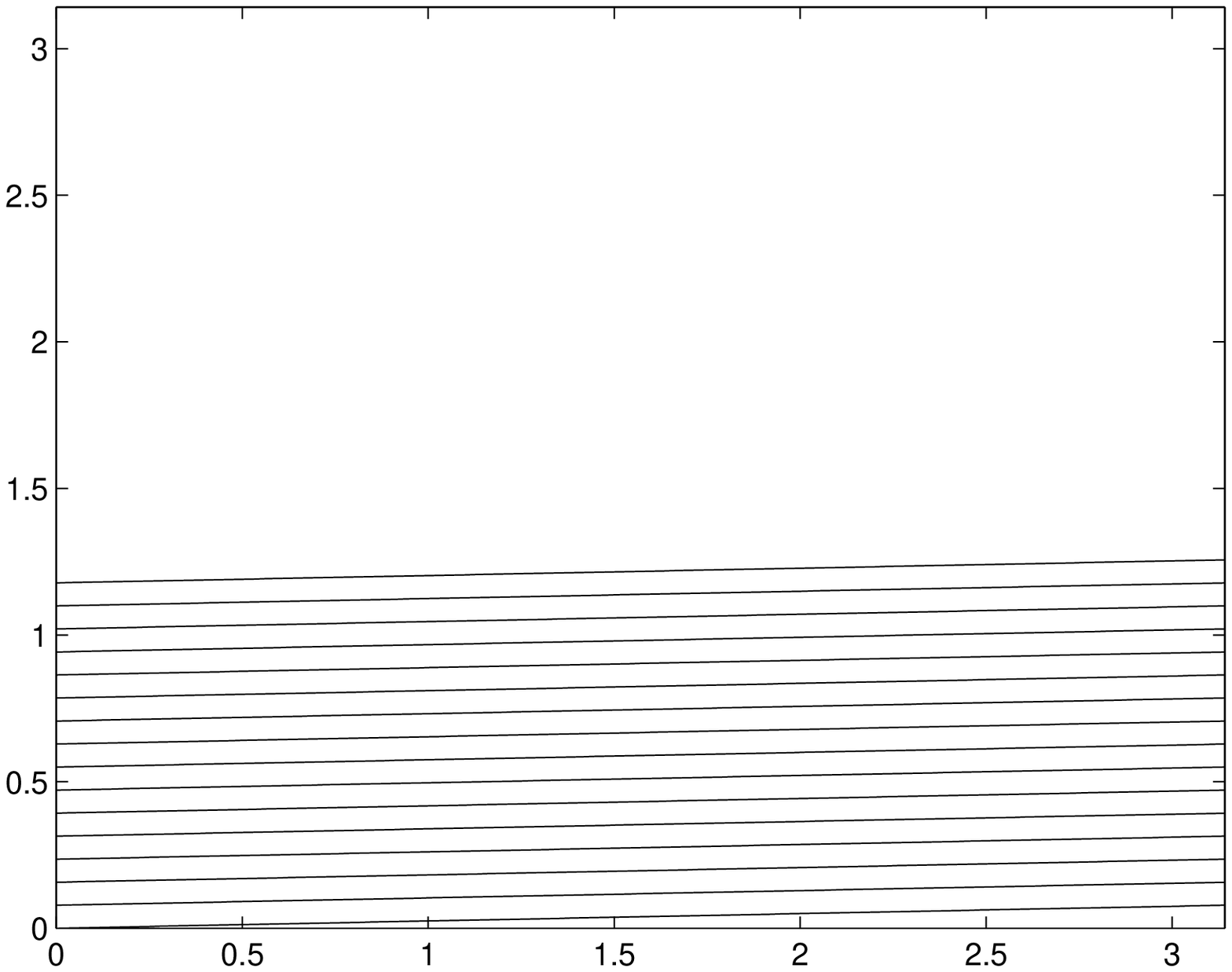,height=160pt,width=160pt}}
\put(80,-8){$\Phi_2$}
\put(-10,80){\rotatebox{90}{$\Phi_1$}}
\end{picture}
\caption{Example of how the line made up by $\Phi_1$ and $\Phi_2$ as
functions of $L/E$ may look like.}
\label{phiexample}
\end{figure}

To see which areas of the plots that are relevant, we look at the definition
of  $\Phi_1$ and $\Phi_2$ in Eq. (\ref{phi1phi2}). The two mass
squared differences have
fixed physical values (although
largely unknown at present).  The quantity $L/E$ can be used as a parameter
for $\Phi_1$ and
$\Phi_2$. Varying $L/E$ corresponds to a line in the plots in Figs.
\ref{p11phiplot}-\ref{p33phiplot} modulo $\pi$. An example of how the
line may look like is given in Fig. \ref{phiexample}.

The slope of the line depends on the ratio between the two mass squared differences
and the size of the area it covers depends on what values $L/E$ can take.
We see that the slope of the line is of little importance when we check if it
intersects a marked area in Figs.
\ref{p11phiplot}-\ref{p33phiplot}. This means that, even when making assumptions of
the mass squared differences to simplify the analysis, it is necessary
to allow the mixing angles to be in $[0,\pi/2]$.

When considering a line as in Fig. \ref{phiexample}, one should
remember that the length of the line depends on the range of values
that $L/E$ can take in an experiment. In a particular experiment this
may be a very short linesegment, but this does not imply that we only have to
consider this short line. By making the experiment better, or using
other experiments measuring the same probability, the line
will be longer and thus any imaginable extension of the experiment
should be taken into account when considering the line in Fig.
\ref{phiexample}. This means that in practice the line will always
cover the whole plot.

The analysis above showed that it is possible to diminish the domain of either
$\theta_1$ or $\theta_2$ without getting any difference of the ranges
of the $P_{\alpha \alpha}$'s. It is, however, not at all clear which of
them one should choose to diminish. As different choices can lead
to different solutions, both of them should be left to be in the
domain $[0,\pi/2]$.

It is possible to show, using the same technique as here, that as
soon as one tries to diminish the domain of variation of all three mixing
angles from $[0,\pi/2]$, it makes a difference in the ranges.

In conclusion, when using a model for three flavor oscillations
as described briefly above, it is necessary and sufficient
to allow all of the mixing angles to be in the domain  $[0,\pi/2]$ when fitting
experimental data, even when making assumptions on the mass squared differences.

\acknowledgments
This work was supported by the Swedish Natural Science Research
Council (NFR), Contract No.  F-AA/FU03281-312.  We want to thank T.
Ohlsson and C. Meier for useful discussions.

\end{document}